\documentclass[12pt,prd,aps,amssymb,amsmath,tightenlines,showpacs,a4paper]{revtex4}

\usepackage{epsfig}
\usepackage{graphicx}
\usepackage{amsmath}
\usepackage{amsfonts,amssymb}
\usepackage{color}

  \usepackage{slashed}
  \usepackage{url}
  \usepackage{amsthm}
\usepackage{float}
\usepackage{bm}
\usepackage{verbatim}

\begin{document}


\title{Exact solutions and zero modes in scalar field theory}


\author{Marco Frasca}
\email[e-mail:]{marcofrasca@mclink.it}
\affiliation{Via Erasmo Gattamelata, 3 \\
             00176 Roma, Italy}


\date{\today}

\begin{abstract}
We provide a set of exact solutions in field theory of scalar fields with $Z_2$ symmetry that involve Jacobi elliptic functions. These solutions have the interesting property to provide massive waves even if one starts from a massless equation. We analyze them classically providing also exact solutions to the corresponding equations for the Green functions needed to completely solve them in a strong coupling limit. This is accomplished using a functional expansion into powers of the current. It is shown that the spontaneous breaking of the $Z_2$ symmetry is due to the existence of a zero mode that persists also in the case of the Higgs mechanism. In this latter case, the zero mode seems to play a role similar to the Goldstone boson in the breaking of a continuous symmetry and so, it should be important at lower momenta as a long range excitation.
\end{abstract}


\maketitle


\section{Introduction}

Exact solutions are generally quite difficult to find but having one for a well-known theory can be really illuminating. A typical theory that, after the recent discovery, proves to be fundamental both in physics and mathematics is that of a scalar field with a nonlinear interaction. A question one could ask is if, choosing a massless theory, it could acquire a mass gap. Generally, this question is rather difficult to answer due to the fact that just perturbation techniques are known to approach this problem and a mass gap could a non-perturbative effect. Indeed, it is quite common to perform lattice studies to understand the behavior of these theories in such regimes (\cite{Cea:1999kn,Cea:2004ka,Balog:2004zd,Stevenson:2005yn,Balog:2006fs} and Refs. therein).  But also in this case, The situation will probably remain unclear also because, due to its non-local nature, the code used for the simulations (cluster algorithm) is not trivially parallelizable. This prevents a straightforward extension to those much larger lattices that, in principle, should be available with the present computer technology. This makes more compelling the availability of theoretical results and so, the knowledge of exact solutions.

Working with these aims in mind, we have published several papers providing exact classical solutions to the scalar field theory \cite{Frasca:2007en,Frasca:2009bc,Frasca:2012ne}. These solutions provide the basis to build up a quantum field theory in a strongly coupled regime \cite{Frasca:2013tma}. Besides, the peculiar property they have is to be massive even if the theory we started from is massless. This mas gap, as we will show in this paper, arises due to the presence of a zero mode. This mode, in the case of a Higgs model, play the role of a would-be Goldstone particle. Jacobi elliptic functions play a fundamental role in all these theories both classically and in the quantum field theory.

This paper has the aim to give a wealth of exact results, based on elliptic functions, for these scalar theories that were never published before. We also yields some differential equations where the Jacobi elliptic functions play the role of a time varying frequency. The cases we consider are all exactly solvable and we also give for them eigenvalues and eigenfunctions. This widens the possible range of applicability of these results.

The paper is so structured. In Sec.\ref{sec:gen} we present the general theory that will be applied in the following sections. In Sec.\ref{sec:mass} we discuss the massive case with the limiting case of the mass going to zero. In Sec.\ref{sec:breaking} we analyze the case of the spontaneous symmetry breaking. In Sec.\ref{sec:spectral}, we prove the existence of the zero mode both in the massless and the Higgs case. Finally, in Sec.\ref{sec:conc}, conclusions are given.

\section{General theory}
\label{sec:gen}

We suppose to have a field theory described by the equation of motion
\begin{equation}
\label{eq:mot0}
   -\Box\phi + \lambda V'(\phi) = j
\end{equation}
being $\Box=-\partial_t^2+\Delta$ the wave operator and $V(\phi)$ the potential with the coupling $\lambda$. This equation has $Z_2$ symmetry given by the transformations $\phi\rightarrow -\phi$ and $j\rightarrow -j$. We assume that the solution to this equation is a functional $\phi=\phi[j]$. Then, we suppose to be able to solve exactly the homogeneous equation
\begin{equation}
   -\Box\phi_0 + \lambda V'(\phi_0) = 0
\end{equation}
where $\phi_0=\phi[0]$. Given this solution, one can build a strongly coupled theory for the
case with $j\ne 0$. In order to do this, one rescales the space-time
variable into eq.(\ref{eq:mot0}) as $x\rightarrow\sqrt{\lambda}x$ and
makes explicit the $\lambda-$dependence of the source as
$j\rightarrow\sqrt{\lambda}j$. Then, one takes the power series
\begin{equation}
 \phi(x)=\sum_{n=0}^\infty\lambda^{-n/2}\phi_n(x)
\end{equation}
and replaces into the equation of motion. A set of non-trivial
equations for $\phi_n$ is obtained whose solution coincides with the
eventual choice to take $\phi$ as a functional of $j$ and do a
functional power series in the current
\cite{Frasca:2012ne,Frasca:2013tma}, after comparison order by order. So, one has
\begin{equation}
\label{eq:jser}
   \phi[j]=\phi_0(x)+\sum_{n=1}^\infty\frac{1}{n!}\int dx_1\ldots dx_n
   \left.\frac{\delta^n\phi}{\delta j(x_1)\ldots\delta j(x_n)}\right|_{j=0}j(x_1)\ldots j(x_n)
\end{equation}
and a set of classical n-point function can be immediately identified
\begin{equation}
   G_n(x_1,\ldots,x_n)=\left.\frac{\delta^n\phi}{\delta j(x_1)\ldots\delta j(x_n)}\right|_{j=0}.
\end{equation}
The equations to get them and completely solve eq.(\ref{eq:mot0}) are easily obtained by successive derivation of this equation. We just note that
\begin{equation}
    -\Box G_1 + V''(\phi_0) G_1 = \delta^4(x)
\end{equation}
and so, $G_1$ is the Green function and all the other n-point functions can be obtained from this \cite{Frasca:2012ne,Frasca:2013tma}. That means that, for a complete solution of the theory, we must be able to solve this equation for the Green function.

This current expansion can be used to study the corresponding quantum field theory in a strong coupling limit. That is, given the generating functional
\begin{equation}
    Z[j]={\cal N}\int[d\phi]\exp\left[i\int d^4x\left(\frac{1}{2}(\partial\phi)^2-V(\phi)+j\phi\right)\right]
\end{equation}
we can immediately substitute the functional series (\ref{eq:jser}) into it and perform computations. So, the leading order has always the form
\begin{equation}
\label{eq:qft0}
    Z_0[j]={\cal N}_0\exp\left[i\int d^4xj\phi_0+\frac{i}{2}\int d^4xd^4x'j(x)\Delta(x-x')j(x')\right]
\end{equation}
where we put $\Delta(x-x')=G_1(x-x')$ and translational invariance will be proved in the cases we consider {\sl a posteriori}. This generating functional is a Gaussian one and all this class of theories is infrared trivial in four dimensions. The linearized theory is represented through the operator
\begin{equation}
  L_{\phi_0}=-\Box+V''(\phi_0)
\end{equation}
and this operator could have a zero mode that breaks spontaneously both the translational and $Z_2$ symmetries of the theory. In this way, a massless theory can acquire a mass while, in the Higgs mechanism breaking $Z_2$ symmetry, this zero mode acts much like a would-be Goldstone boson.

\section{Massive scalar field theories}
\label{sec:mass}

Let us consider the equation
\begin{equation}
-\Box\phi +\mu_0^2\phi+\lambda\phi^3 = j
\end{equation}
being $\mu_0$ the mass of the field and $\lambda$ the coupling. By direct substitution one can verify that a solution to the homogeneous equations is given by
\begin{equation}
\phi_0(x) = \pm\mu\left(\frac{2}{\lambda}\right)^\frac{1}{4}{\rm sn}
\left(p\cdot x+\theta,i\sqrt{\frac{\sqrt{2\lambda}\mu^2}{2\mu_0^2+\sqrt{2\lambda}\mu^2}}\right)
\end{equation}
provided that
\begin{equation}
   p^2=\mu_0^2+\mu^2\sqrt{\frac{\lambda}{2}}
\end{equation}
being $\theta$ and $\mu$ two integration constants and ${\rm sn}$ a Jacobi elliptic function that is periodic representing a nonlinear wave-like solution. From the dispersion relation we can see that the mass of the field gets a contribution from the coupling $\lambda$ and the integration constant $\mu$, so it arises from the self-interacting term as it becomes zero when $\lambda$ is taken to be zero (no interaction case).

These formulas are simplified setting the mass of the field $\mu_0$ to zero. The solution becomes
\begin{equation}
\label{eq:phis}
   \phi_0(x) = \pm\mu\left(\frac{2}{\lambda}\right)^\frac{1}{4}{\rm sn}(p\cdot x+\theta,i) 
\end{equation}
provided
\begin{equation}
   p^2=\mu^2\left(\frac{\lambda}{2}\right)^\frac{1}{2}.
\end{equation}
This solution shows how a massless field can become massive just from the self-interacting term. Already at the classical level, we get an arbitrary integration constant having the dimension of a mass. We will see below that, for the most common scalar theories, we are able to solve them completely in the limit of the coupling $\lambda$ running to infinity. This will yield a dual power series to the standard weakly coupled expansion in perturbation theory.

Back to the massive case, we want to completely solve this case and we need to find a solution to the equation for the Green function
\begin{equation}
    -\Box\Delta(x) + [\mu_0^2+3\lambda\phi_0^2(x)]\Delta(x) = \delta^4(x)
\end{equation}
that we rewrite as
\begin{equation}
    -\Box\Delta(x) + \left[\mu_0^2+3\mu^2\left(2\lambda\right)^\frac{1}{2}{\rm sn}^2
\left(p\cdot x+\theta,
i\sqrt{\frac{\sqrt{2\lambda}\mu^2}{2\mu_0^2+\sqrt{2\lambda}\mu^2}}\right)\right]\Delta(x)=\delta^4(x).
\end{equation}
In order to solve this equation, we consider the rest frame putting ${\bm p}=0$. Then, we set $\Delta(x)=\delta^3(x)\bar\Delta(t)$ and we have to solve
\begin{equation}
    -\partial_t^2\bar\Delta(t) + \left[\mu_0^2+3\mu^2\left(2\lambda\right)^\frac{1}{2}{\rm sn}^2
\left(mt+\theta,i\sqrt{\frac{\sqrt{2\lambda}\mu^2}{2\mu_0^2+\sqrt{2\lambda}\mu^2}}\right)\right]\bar\Delta(t)=\delta(t).
\end{equation}
where we have introduced the effective mass of the field $m=\sqrt{\mu_0^2+\mu^2\left(\frac{\lambda}{2}\right)^\frac{1}{2}}$. This represents the first of a kind of linear differential equations describing a harmonic oscillator with a time varying frequency and the variation law is represented using Jacobi elliptic functions. This kind of equations can be exactly solved as we show in a moment. The solution can be immediately written down as
\begin{eqnarray}
    \bar\Delta_n(t)&=&-\theta(t)Z_\Delta
   {\rm cn}\left[m t+\theta_n,
   i\sqrt{\frac{\sqrt{2\lambda}\mu^2}{2\mu_0^2+\sqrt{2\lambda}\mu^2}}\right]\times \nonumber \\
   &&{\rm dn}\left[m t+\theta_n,
   i\sqrt{\frac{\sqrt{2\lambda}\mu^2}{2\mu_0^2+\sqrt{2\lambda}\mu^2}}\right]
\end{eqnarray}
having fixed the phases to $\theta_n=(4n+1)K\left(i\sqrt{\frac{\sqrt{2\lambda}\mu^2}{2\mu_0^2+\sqrt{2\lambda}\mu^2}}\right)$ with $n=0,\pm 1,\pm 2,\ldots$, 
${\rm cn}$ and ${\rm dn}$ being Jacobi elliptic functions, $K(\alpha)=\int_0^\frac{\pi}{2}d\theta/\sqrt{1-\alpha^2\sin^2\theta}$ the elliptic integral of the first kind, and
\begin{equation}
Z_\Delta=\frac{(2m^2)^\frac{7}{2}}{4\left(2^\frac{1}{2}9\mu_0^4\mu^4\lambda +8^\frac{1}{2}\mu_0^8+10\mu_0^6\mu^2\lambda^\frac{1}{2}+2^\frac{1}{2}\mu^8\lambda^2+7\mu_0^2\mu^6\lambda^\frac{3}{2}\right)}.
\end{equation}
This identifies an infinite set of infrared trivial scalar field theories.  Then, we notice that
\begin{equation}
    \bar\Delta_n(t)=-\theta(t)Z_\Delta\left.\frac{d}{du}{\rm sn}
    \left(u,i\sqrt{\frac{\sqrt{2\lambda}\mu^2}{2\mu_0^2+\sqrt{2\lambda}\mu^2}}\right)
    \right|_{u=m t+(4n+1)K\left(i\sqrt{\frac{\sqrt{2\lambda}\mu^2}{2\mu_0^2+\sqrt{2\lambda}\mu^2}}\right)}
\end{equation}
and recall that
\begin{equation}
    \operatorname{sn}(u,ik)=\frac{2\pi}{K(ik)}\sum_{n=0}^\infty \frac{q^{n+1/2}}{1-q^{2n+1}} \sin\left[(2n+1)\frac{\pi u}{2K(ik)}\right]
\end{equation}
being $q=\exp\left(-\pi K'(ik)/K(ik)\right)$ with $K'(ik)=K(\sqrt{1+k^2})$ as in our case $k=\sqrt{\frac{\sqrt{2\lambda}\mu^2}{2\mu_0^2+\sqrt{2\lambda}\mu^2}}$ that is $k=1$ when $\mu_0=0$, the massless limit of the theory. This will yield
\begin{equation}
    \bar\Delta(t)=\theta(t)Z_\Delta
    \frac{2\pi^2}{K^2(ik)}\sum_{r=0}^\infty (-1)^n(2r+1)\frac{q^{r+1/2}}{1-q^{2r+1}}\sin\left[(2r+1)\frac{\pi}{2K(ik)}m t\right]
\end{equation}
where we have fixed the phase without losing generality.  We can now go back to the full propagator and take the Fourier transform obtaining
\begin{equation}
    \Delta(\omega,0)=mZ_\Delta\frac{2\pi^3}{iK^3(ik)}\sum_{r=0}^\infty (-1)^n(2r+1)^2\frac{q^{r+1/2}}{1-q^{2r+1}}\frac{1}{\omega^2-m_n^2+i\epsilon}.
\end{equation}
Finally, after boosting the rest frame, one has the full propagator
\begin{equation}
\label{eq:prop1}
    \Delta(p)=mZ_\Delta\frac{2\pi^3}{iK^3(ik)}\sum_{r=0}^\infty (-1)^r(2r+1)^2\frac{q^{r+1/2}}{1-q^{2r+1}}\frac{1}{p^2-m_r^2+i\epsilon}.
\end{equation}
being 
\begin{equation}
\label{eq:ms1}
    m_n=(2n+1)\frac{\pi}{2K(ik)}m.
\end{equation}
This solves completely the classical theory for a massive scalar field theory in the limit of the coupling $\lambda$ running to infinity. Quantum field theory at the leading order can be written down immediately using eq.(\ref{eq:qft0}). The theory is infrared trivial with a mass spectrum given by eq.(\ref{eq:ms1}). This describes a free particle with a harmonic oscillator spectrum superimposed. We emphasize that {\sl translational invariance is shown a posteriori in the propagator} but this symmetry is spontaneously broken by zero modes as we will show in Sec.\ref{sec:spectral}. The propagator displays translational invariance but broken $Z_2$ symmetry depending on the choice of the $\mu$ parameter.

For the sake of completeness, we give below the same results for the massless case ($\mu_0=0$). One has for the Green function
\begin{equation}
    -\Box\Delta_{\mu_0^2=0}(x) + 3\lambda\phi_0^2(x)\Delta_{\mu_0^2=0}(x) = \delta^4(x)
\end{equation}
that we rewrite as
\begin{equation}
    -\Box\Delta_{\mu_0^2=0}(x) + 3\mu^2\left(2\lambda\right)^\frac{1}{2}{\rm sn}^2
     \left(p\cdot x+\theta,i\right)\Delta_{\mu_0^2=0}(x)=\delta^4(x).
\end{equation}
The solution of this equation can be immediately written down from eq.(\ref{eq:prop1}) and yields
\begin{equation}
\label{eq:prop2}
    \Delta_{\mu_0^2=0}(p)=\frac{\pi^3}{4K^3(i)}\sum_{r=0}^\infty(2r+1)^2\frac{e^{-(r+\frac{1}{2})\pi}}{1+e^{-(2r+1)\pi}}\frac{1}{p^2-m_r^2+i\epsilon}
\end{equation}
being
\begin{equation}
    K(i)=\int_0^{\pi/2}{\frac{dx}{\sqrt{1+\sin^2x}}}=1.3110287\ldots
\end{equation}
the complete elliptic integral of first kind and provided the mass spectrum is 
\begin{equation}
\label{eq:ms2}
    m_n=(2n+1)\frac{\pi}{2K(i)}\left(\frac{\lambda}{2}\right)^\frac{1}{4}\mu.
\end{equation}
This shows that this class of solutions for the scalar field implies a mass gap into the quantum theory even if we started from a massless field. These results agree perfectly well and extend those given in Ref.\cite{Frasca:2009bc,Frasca:2013tma}.

Finally, we note that this representation can be seen
as a form of K\"allen-Lehmann decomposition for the propagator
\begin{equation}
    \Delta(p)=\int_0^\infty ds\frac{\rho(s)}{p^2-s+i\epsilon}
\end{equation}
with a spectral function
\begin{equation}
     \rho(s)= \sum_{r=0}^\infty\delta(s-m^2_r)\rho_r
\end{equation}
and
\begin{equation}~
\rho_r=mZ_\Delta\frac{2\pi^3}{iK^3(ik)}(-1)^r(2r+1)^2\frac{q^{r+1/2}}{1-q^{2r+1}}\qquad {r=0,1,2...}
\end{equation}
This spectral function, which corresponds to a sum of Yukawa
propagators without bound states, when taken as model of the
continuum limit for the corresponding quantum field theory, provides
a good representation of ``triviality''. In fact, the ratios
$\rho_r/\rho_0$ vanish very rapidly with increasing $k$ so that the
properly normalized strength of the single-particle pole is largely prevailing.
The theory describes free particles with a superimposed harmonic oscillator spectrum.

\section{Spontaneous breaking of symmetry}
\label{sec:breaking}

We get spontaneous breaking of symmetry when the mass term is taken with a wrong sign to have
\begin{equation}
\label{eq:mot2}
   -\Box\phi -\mu_0^2\phi +\lambda\phi^3= j.
\end{equation}
This equation has the following exact wave-like solution for $j=0$
\begin{equation}
   \phi_{SSB}(x) =\pm v\cdot {\rm dn}(p\cdot x+\theta,i)
\end{equation}
provided that
\begin{equation}
   p^2=\frac{\lambda v^2}{2}
\end{equation}
being $v=\sqrt{\frac{2\mu_0^2}{3\lambda}}$. We see that the dispersion relation has the mass term with the right sign and we are describing oscillations around one of the selected solutions $\phi=\pm \sqrt{\frac{3}{2}}v$ as the Jacobi function ${\rm dn}$ is never zero. Looking at the Fourier series for this solution (\ref{eq:foussb}), we note that there is a contribution coming form a zero frequency mode. This mode will contribute in a quantum field theory and must be retained. We will clarify its role below. 

Assuming as done above that $\phi=\phi[j]$ and performing the functional derivative on the equation of motion as required in (\ref{eq:jser}), one recovers the Green function as the first functional derivative of the eq. (\ref{eq:mot2}). This gives
\begin{equation}
   \partial^2\Delta(x)+\mu_0^2\left[2\cdot{\rm dn}^2(p\cdot x+\theta,i)-1\right]\Delta(x)=\delta^4(x)
\end{equation}
In order to solve this equation, we follow the procedure given in
in the preceding section. We consider the rest frame where this equation
can be cast in the form
\begin{equation}
   \partial^2_t\bar\Delta(t)+\mu_0^2\left[2\cdot{\rm dn}^2\left(\frac{\mu}{\sqrt{3}}t+\theta,i\right)-1\right]\bar\Delta(t)=\delta(t)
\end{equation}
and we consider an overall multiplying factor $\delta^3({\bf x})$ in
the final solution. This is due to the linearity of the equation.
This equation has the class of solutions (labeled by an integer $n$)
\begin{equation}
   \Delta_n({\bf x},t)=\delta^3({\bf x})\bar\Delta(t)=-\delta^3({\bf x})\theta(t)
   \sqrt{\frac{3}{2}}\frac{1}{\mu_0}{\rm sn}\left(\frac{\mu_0}{\sqrt{3}}t+\theta_n,i\right)
   {\rm cn}\left(\frac{\mu_0}{\sqrt{3}}t+\theta_n,i\right)
\end{equation}
where ${\rm sn}(u,i)$ and ${\rm cn}(u,i)$ are elliptic functions, $\theta_n=(2n+1)K(i)$ are phases that must be properly fixed. We also note that
\begin{equation}
   \Delta_n({\bf x},t)=-\delta^3({\bf x})\theta(t)\sqrt{\frac{3}{2}}\frac{1}{\mu_0}
   \left.\frac{d}{du}{\rm dn}(u,i)\right|_{u=\frac{\mu}{\sqrt{3}}t+\theta_n}.
\end{equation}
Now, we observe that
\begin{equation}
    {\rm dn}(u,i)=-\frac{\pi}{2K(i)}+\frac{2\pi}{K(i)}\sum_{k=0}^\infty(-1)^k\frac{e^{-k\pi}}{1+e^{-2k\pi}}
    \cos\left(\frac{2\pi ku}{2K(i)}\right).
\end{equation}
in order to keep the zero-frequency mode, one gets
\begin{equation}
   \Delta_n({\bf x},t)=\delta^3({\bf x})\theta(t)\frac{\sqrt{3}}{\mu_0}
   \frac{\sqrt{2}\pi^2}{K^2(i)}\sum_{k=0}^\infty(-1)^kk\frac{e^{-k\pi}}{1+e^{-2k\pi}}
    \sin\left(\frac{2\pi k}{2K(i)}\left(\frac{\mu_0}{\sqrt{3}}t+\theta_n\right)\right).
\end{equation}
The term with $k=0$ formally vanishes and appears to be non-propagating (except for a possible ambiguity in the limit $p\rightarrow 0$ \cite{Consoli:2002ha,Consoli:2006ji,Consoli:2009rp,Zappala:2012wh}. Then, following the procedure of the preceding section we have
\begin{equation}
   \Delta({\bf p}=0,\omega)=\frac{\sqrt{3}}{\mu_0}
   \frac{\sqrt{2}\pi^2}{K^2(i)}
   \sum_{k=0}^\infty k\frac{e^{-k\pi}}{1+e^{-2k\pi}}
   \frac{m_k}{\omega^2-m_k^2+i\epsilon}
\end{equation}
with
\begin{equation}
\label{eq:spec}
   m_k=k\frac{\pi}{K(i)}\frac{\mu_0}{\sqrt{3}}\qquad (k=0,1,2,\ldots).
\end{equation}
In this way, by boosting to a moving Lorentz frame, we get the
propagator in the classical theory
\begin{equation}
\label{eq:prop3}
    \Delta(p)=\sqrt{2}\frac{\pi^3}{K^3(i)}
    \sum_{k=0}^\infty
    k^2\frac{e^{-k\pi}}{1+e^{-2k\pi}}\frac{1}{p^2-m_k^2+i\epsilon}
\end{equation}
with the poles given in eq.(\ref{eq:spec}). These poles and the form
of the propagator confirm the results given in
\cite{Frasca:2007en,Frasca:2009bc}. We have explicitly kept the ambiguity at zero momenta of the propagator as we have got this result choosing a particular order in the limit procedures that are involved here.

Following the analysis for quantum field theory, we can immediately write down the generating functional in the infrared limit using again eq.(\ref{eq:qft0}). The theory is infrared trivial with a zero mass mode that left an open question about its nature of propagating or non-propagating mode due to the presence of a non-commuting double limit operation as already pointed out in literature \cite{Consoli:2002ha,Consoli:2006ji,Consoli:2009rp,Zappala:2012wh}.

\section{Spectral properties of the quantum field theory}
\label{sec:spectral}

In this section we will consider the cases of the massless scalar field and spontaneous breaking of symmetry. We observe that a massless field becomes massive and the only way this can happen is by spontaneous symmetry breaking. So, we should understand the origin of the mechanism of it. The mechanism acting here is identical to that described in Ref.\cite{Shifman:2012zz} and is due to the presence of a zero mode. This can be seen in the following way. The Hamiltonian of the system is given by
\begin{equation}
    H=\int d^3x\left[\frac{1}{2}(\partial_t\phi)^2+\frac{1}{2}(\nabla\phi)^2+\frac{\lambda}{4}\phi^4\right].
\end{equation} 
We linearize it around the classical solution (\ref{eq:phis})
\begin{equation}
    \phi(x)=\phi_0(x)+\delta\phi(x)
\end{equation}
yielding
\begin{equation}
   H=H_0+\int d^3x\left[\frac{1}{2}(\partial_t\delta\phi)^2+\frac{1}{2}(\nabla\delta\phi)^2
   +\frac{3}{2}\lambda\phi_0^2\delta\phi^2\right]+O\left(\delta\phi^3\right)
\end{equation}
being $H_0$ the contribution coming from the classical solution. The linear part can be diagonalized with a Fourier series provided we are able to get the eigenvalues and the eigenvectors of the operator
\begin{equation}
   L_{\mu_0^2=0}=-\Box+3\lambda\phi_0^2(x).
\end{equation}
It is not difficult to realize that there is a zero mode. We give the solutions for both the zero and non-zero modes. The spectrum is continuous with eigenvalues 0 and $3\mu^2\sqrt{\lambda/2}$ with $\mu$ varying continuously from 0 to infinity. The zero mode solution has the aspect
\begin{equation}
  \chi_0(x,\mu)=a_0\cdot{\rm cn}(p\cdot x+\theta,i)\cdot{\rm dn}(p\cdot x+\theta,i)
\end{equation}
being $a_0$ a normalization constant. Non-zero modes are given by
\begin{equation}
  \chi(x,\mu)=a'\cdot{\rm sn}(p\cdot x+\theta,i)\cdot{\rm dn}(p\cdot x+\theta,i).
\end{equation}
with $a'$ again a normalization constant. These hold on-shell, that is when $p^2=\mu^2\sqrt{\lambda/2}$. The spectrum is continuous and so, these eigenfunctions are not normalizable. So, we note that there is a doubly degenerate set of zero modes spontaneously breaking translational invariance and the $Z_2$ symmetry of the theory. As seen in Sec.\ref{sec:mass}, the phases $\theta$ are fixed to the values $(4n+1)K(i)$ in the massless case. This yields for the zero modes
\begin{equation}
  \chi_0(x,\mu)=-2a_0\frac{{\rm sn}(p\cdot x,i)}{{\rm dn}^2(p\cdot x,i)}.
\end{equation}
For a given $\mu$ parameter, $Z_2$ symmetry is spontaneously broken through this zero mode. This mode disappears when $\mu=0$ as it should.

In the case of breaking of symmetry, the situation is identical. One has for the eigenfunctions and eigenvalues of the operator
\begin{equation}
   L_{\mu_0^2\ne 0}=-\Box-\mu_0^2+3\lambda\phi_{SSB}^2(x).
\end{equation}
Also in this case it is not difficult to realize that there is a zero mode. The spectrum is continuous with eigenvalues 0 and $\mu_0^2$ with $\mu_0$ varying continuously in $(0,\infty)$. The zero mode solution has the aspect
\begin{equation}
  \bar\chi_0(x,\mu_0)=b_0\cdot{\rm sn}(p\cdot x+\theta,i)\cdot{\rm cn}(p\cdot x+\theta,i)
\end{equation}
being $b_0$ a normalization constant. Non-zero modes are given by
\begin{equation}
  \bar\chi(x,\mu_0)=b'\cdot{\rm cn}(p\cdot x+\theta,i)\cdot{\rm dn}(p\cdot x+\theta,i).
\end{equation}
with $b'$ again a normalization constant. These hold on-shell with $p^2=\frac{\lambda v^2}{2}$. Again, the spectrum is continuous and the eigenfunctions are not normalizable. So, we note that there is a doubly degenerate set of zero modes spontaneously breaking translational invariance and the $Z_2$ symmetry of the theory. As seen in Sec.\ref{sec:mass}, the phases $\theta$ are fixed to the values $(2n+1)K(i)$ in the massless case. This yields for the zero modes
\begin{equation}
  \bar\chi_{0}(x,\mu_0)=-2b_0\frac{{\rm sn}(p\cdot x,i)\cdot{\rm cn}(p\cdot x,i)}{{\rm dn}^2(p\cdot x,i)}.
\end{equation}
For a given $\mu_0$ parameter, $Z_2$ symmetry is spontaneously broken but this zero mode can play a role in the infrared limit similar to a Goldstone boson in the breaking of a continuous symmetry. This mode cannot be set to zero in this case as the theory with $\mu_0=0$ becomes the massless one. In this way it is seen that these theories are connected each other continuously. 

\section{Conclusions}
\label{sec:conc}

We have shown how a scalar field theory with $Z_2$ symmetry can be exactly solved and conclusions can be drawn both classically and for the quantum theory. These solutions are interesting for a couple of reasons. Mathematically, because there is a large usage of Jacobi elliptic functions that appear to behave like some kind of nonlinear plane waves. Physically, because we see a zero mode that breaks $Z_2$ symmetry spontaneously and acts like a kind of Goldstone boson in a Higgs mechanism. Last but not least, the overall effect is to see massless field to become massive.

There is a possibility that these solutions could be chosen in nature to describe the way mass appears at fundamental level. This appears like another possibility that could find application anyway in higher level theories beyond the Standard Model.



\section*{Acknowledgements}

I would like to thank the organizers of the Workshop
on QCD in strong magnetic fields at Catania University, Vincenzo Greco and Marco Ruggieri,
where the idea for this work was firstly born. Interesting discussions with Maurizio Consoli were also instrumental to complete this paper.

\appendix*
\section{Fourier expansion of solutions}
\label{sec:fourier}

The exact solutions discussed in this paper have a Fourier expansion, being periodic functions. These expansions are widely known, being those of Jacobi elliptic functions. In this way one can present them as a superposition of plane waves. So, one has \cite{gra}
\begin{equation}
   \phi(x) = \frac{2\pi}{kK(k)}\sum_{n=0}^\infty\frac{q^{n+\frac{1}{2}}}{1-q^{2n+1}}
   \sin\left[(2n+1)\frac{\pi}{2K(k)}p\cdot x\right]
\end{equation}
being $q=e^{-\pi\frac{K'(k)}{K(k)}}$, 
$k=i\sqrt{\frac{\sqrt{2\lambda}\mu^2}{2\mu_0^2+\sqrt{2\lambda}\mu^2}}$, $K'(k)=K(\sqrt{1-k^2})$ and finally $K(k)=\int_0^{\pi/2}dx/\sqrt{1-k^2\sin x}$. The reason why such a Fourier series is interesting is that, in the rest reference frame, taking ${\bf p}=0$, one has
\begin{equation}
   \phi(t,0) = \frac{2\pi}{kK(k)}\sum_{n=0}^\infty\frac{q^{n+\frac{1}{2}}}{1-q^{2n+1}}
   \sin\left[(2n+1)\frac{\pi}{2K(k)}mt\right]
\end{equation} 
being
\begin{equation}
   m=m(\mu_0,\mu,\lambda)=\sqrt{\mu_0^2+\mu^2\left(\frac{\lambda}{2}\right)^\frac{1}{2}}
\end{equation}
that is the ``renormalized mass'' for these classical field excitations. From this expansion one can read out a kind of mass spectrum
\begin{equation}
   \epsilon_n=(2n+1)\frac{\pi}{2K(k)}m.
\end{equation}
These equations simplify significantly taking $\mu_0=0$.One has
\begin{equation}
   \phi(x)=\pm\mu\left(\frac{2}{\lambda}\right)^\frac{1}{4}\sum_{n=0}^\infty(-1)^n\frac{2\pi}{K(i)}
   \frac{e^{-\left(n+\frac{1}{2}\right)\pi}}
   {1+e^{-(2n+1)\pi}}\sin\left((2n+1)\frac{\pi}{2K(i)}p\cdot x\right)
\end{equation}
so that
\begin{equation}
\label{eq:ser}
   \phi(t,0)=\pm\mu\left(\frac{2}{\lambda}\right)^\frac{1}{4}\sum_{n=0}^\infty(-1)^n\frac{2\pi}{K(i)}
   \frac{e^{-\left(n+\frac{1}{2}\right)\pi}}
   {1+e^{-(2n+1)\pi}}\sin\left((2n+1)\frac{\pi}{2K(i)}\left(\frac{\lambda}{2}\right)^\frac{1}{4}\mu t\right)
\end{equation}
and a ``mass spectrum''
\begin{equation}
   \epsilon_n = (2n+1)\frac{\pi}{2K(i)}\left(\frac{\lambda}{2}\right)^\frac{1}{4}\mu.
\end{equation}
Finally, we consider the case for spontaneous breaking of symmetry having
\begin{equation}
\label{eq:foussb}
   \phi(x)=\frac{\pi}{2K(i)}+\frac{2\pi}{K(i)}\sum_{n=1}^\infty\frac{e^{-n\pi}}{1+e^{-2n\pi}}
   \cos\left(2n\frac{\pi}{2K(i)}p\cdot x\right)
\end{equation}
and so
\begin{equation}
   \phi(t,0)=\frac{\pi}{2K(i)}+\frac{2\pi}{K(i)}\sum_{n=1}^\infty\frac{e^{-n\pi}}{1+e^{-2n\pi}}
   \cos\left(2n\frac{\pi}{2K(i)}\frac{\mu_0}{\sqrt{3}} t\right)
\end{equation}
giving rise to a ``mass spectrum'' \cite{Frasca:2007en}
\begin{equation}
   \epsilon_n=n\frac{\pi}{K(i)}\frac{\mu_0}{\sqrt{3}}
\end{equation}
with $n=0,1,2,\ldots$.

\end{document}